# Design and Analysis of a Secure Three Factor User Authentication Scheme Using Biometric and Smart Card

Hossen Asiful Mustafa and Hasan Muhammad Kafi

*Abstract*— Password security can no longer provide enough security in the area of remote user authentication. Considering this security drawback, researchers are trying to find solution with multifactor remote user authentication system. Recently, three factor remote user authentication using biometric and smart card has drawn a considerable attention of the researchers. However, most of the current proposed schemes have security flaws. They are vulnerable to attacks like user impersonation attack, server masquerading attack, password guessing attack, insider attack, denial of service attack, forgery attack, etc. Also, most of them are unable to provide mutual authentication, session key agreement and password, or smart card recovery system. Considering these drawbacks, we propose a secure three factor user authentication scheme using biometric and smart card. Through security analysis, we show that our proposed scheme can overcome drawbacks of existing systems and ensure high security in remote user authentication.

## I. Introduction

TODAY, most authentication systems rely on passwords. A password is a word or string of characters used for authentication by a user to prove identity or to gain access approval to a resource (example: an access code is a type of password); a password is to be kept secret from those not allowed access. So, a numerous attempts [1]-[4] have been taken to protect password from being compromised. However, the crackers are able to steal password despite those attempts. This is due to the user's habit of using password. Most of the users use weak passwords [5]; they also reuse same password in several accounts that causes domino effect [6]. According to [5], the average user has 6.5 passwords, each of which is shared across 3.9 different websites. Each user types an average of 8 passwords per day with an average bit-strength of 40.54 bits and has approximate 25 web accounts that require passwords to access. It is also mentioned that a large number of passwords chosen by the users only contain lowercase letters unless forced to do otherwise. Such weak passwords can be guessed by the attacker easily. Increasing password strength may be a solution, but it cannot increase the security by a large degree. The strength of password can be increased by mixing up the lowercase letters, uppercase letters, digits, signs and special characters. But, such kinds of passwords are difficult to remember. It reduces the security of a system because (1) users might need to store these passwords (2) users will need to reset them frequently, and (3) users are more likely reuse the same password. According to [5] at least 1.5% of Yahoo users forget their passwords each month.

For these reasons, several methods like password management software, graphical password schemes, cognitive authentication scheme, one time passwords, hardware tokens, phone aided schemes, biometrics, etc., are proposed to replace password [7]. However, none of them provides good usability, deployability, and strong security at the same time [7]. In 2000 and after wards, several smart card based schemes were proposed [8]-[10]. But, all of them have several weaknesses.

Recently, biometric and smart card based user authentication schemes along with password have drawn a considerable attention from the researchers [11]-[16]. The biometric keys are based on physiological and behavioral characteristics of persons such as fingerprints, faces, irises, hand geometry, palm-prints, etc. The advantages of using biometric key are given bellow:

- They cannot be forgotten or lost.
- They cannot be copied or shared easily.
- They are extremely hard to forge or distribute.
- They cannot be guessed easily.
- They prevent non-repudiation.

Considering these advantages, researchers develop their schemes to enhance the security of remote user authentication system. But, all of their schemes have several weaknesses. In particular, the scheme proposed by Li et al. [16] is vulnerable to attacks like user impersonation attack, server masquerading attack, password guessing attack, insider attack, denial of service attack, forgery attack, etc. The other proposed schemes

Date of submission 3 June 2017.
Dr. Hossen Asiful Mustafa is with the Institute of Information and Communication Technology, Bangladesh University of Engineering and Technology, Dhaka, Bangladesh (e-mail: hossen_mustafa@ iict.buet.ac.bd).
Hasan Muhammad Kafi is with the Institute of Information and Communication Technology, Bangladesh University of Engineering and Technology, Dhaka, Bangladesh (e-mail: engr.kafi@gmail.com).

are also vulnerable to one or more of the above mentioned attacks. Additionally, most of them are unable to provide mutual authentication, session key agreement and password, or smart card recovery system.

In this paper, we propose a three factor user authentication scheme using biometric and smart card that can resist all the above mentioned attacks and is able to provide mutual authentication, session key agreement and password, or smart card recovery system.

The rest of this paper is organized as follows: Section II discusses related works, Section III describes Li et al. [16] scheme briefly and Section IV discusses the security flaws of their scheme. Then, proposed scheme and its security analysis are presented in Section V and Section VI respectively. Finally, we draw our conclusion in Section VII.

## II. RELATED WORKS

A highly secure system requires secure authentication system to deal with high level of security risk. To ensure that type of security, multi-factor user authentication system comes into account. Nowadays, the biometric and smart card based user authentication schemes along with password have drawn a considerable amount of attention from the researchers [11]-[16].

In 2010, an efficient biometric-based remote user authentication scheme using smart card [11] was proposed by Hwang et al. They claimed that the computation cost of their work was relatively low compared with other related schemes. Their proposed scheme is based on smart card, one way hash function and biometric verification. Their scheme can resist masquerading attack, replay attack, parallel session attack and can provide the security of the information stored within the smart card. Moreover, it enables the user to change their password freely, provides mutual authentication between the remote server and the user, doesn't need to store any password or identity tables, doesn't require any synchronized clock and provides non-repudiation.

Das presented an analysis and improvement on an efficient biometric based remote user authentication scheme using smart cards in [14]. He claimed that the scheme proposed by Hwang et al. [11] has design flaws in login and authentication phase, password change phase, and verification of biometrics using hash function. He provided an improved scheme to fix these flaws. In this scheme, the user can freely change password by providing both old and new passwords. He also provided a security analysis to show that their scheme can fix the security flaws of [11].

In [15], An revealed the security flaws of the scheme proposed by Das [14]. He showed that the scheme presented in [14] is vulnerable to user impersonation attack, server masquerading attack, password guessing attack, and insider attack and it cannot provide mutual authentication. He proposed security analysis and enhancements of an effective biometric-based remote user authentication scheme using smart cards to overcome these security weaknesses of [14], while preserving all its merits. He claimed that his scheme can prevent user impersonation attack, server masquerading attack, password guessing attack, insider attack and can provide mutual authentication.

In 2013, Li et al. conducted a detail analysis on [15] and revealed some weaknesses such as vulnerability to denial of service attack, and forgery attack and inability to provide session key agreement [16]; they proposed a robust biometric based remote user authentication scheme with session key agreement using elliptical curve cryptography [16] to overcome these weaknesses. They also provided a security analysis to show that their scheme can overcome the security flaws of the scheme proposed in [15]. They claimed that their scheme can provide the security of secret key, session key agreement, proper biometric authentication, quick detection of unauthorized login, proper mutual authentication, prevent forgery attack, stolen smart card attack, and replay attack. In the following section, we briefly discuss the scheme presented by Li et al.

## III. REVIEW OF LI ET AL.'S SCHEME

The scheme uses a fuzzy extractor which is a pair of procedure $(Gen, Rep)$ such that:

$$Gen(B_i) = (R_i, P_i) \quad (1)$$
$$Rep(B_{ni}, P_i) = R_i \quad (2)$$

where, $B_{ni}$ is reasonably close to $B_i$.

TABLE I
THE NOTATIONS USED THROUGH OUT THIS PAPER

| Notation | Description |
|---|---|
| $R$ | Trusted registration center |
| $S_i$ | Server |
| $C_i$ | User |
| $A_i$ | An attacker |
| $ID_i$ | Identity of the user $C_i$ |
| $PW_i, PW_{ni}$ | Password and new password of the user $C_i$ |
| $B_i, B_{ni}$ | Biometric template or key of the user $C_i$ |
| $R_{cont}$ | Recovery contact of the user $C_i$ |
| $P, n$ | Two large prime numbers |
| $F_p$ | A finite field |
| $E_p(a, b)$ | An elliptic curve defined on finite field $F_p$ with prime number order $n$ |
| $P$ | A point on elliptic curve $E_p(a, b)$ with order $n$ |
| $h(.)$ | A secure hash function |
| $K_s$ | Master secret key for server |
| $X_s$ | Master secret key for user |
| $K, W, R_1$–$R_8$ | A secret random strings |
| $RK_{aes}$ | The AES key of the trusted registration center $R$ |
| $SK_{aes}$ | The AES key of the server $S_i$ |
| $E_{aes}(.)$ | Encryption function for AES |
| $D_{aes}(.)$ | Decryption function for AES |
| $K_{ses}$ | Session key |
| $Stat$ | Status message |
| $\|$ | Message concatenation operator |
| $\oplus$ | Exclusive-or operator |

Initially, $R$ chooses an elliptic curve equation $E_p(a,b)$ over a prime finite field $F_p$ and a base point $P$ with order $n$ over $E_p(a,b)$ and publishes parameters $(E_p(a,b), n, P)$. It also chooses a master secret key $X_s$ and distributes it to the server $S_i$ through a secure channel.

There are four phases in this scheme: the registration phase, the login phase, the authentication and key agreement phase, and the password change phase.

### A. Registration Phase

During this phase, the registration center $R$ and the user $C_i$ have to perform the following steps:

*1) Registration Request*
The user $C_i$ provides his $ID_i$, $PW_i$, $B_i$ at the fuzzy extractor and a random number $K$. The $C_i$ sends $ID_i$, $B_i$, $RPW_i$ to the registration center $R$ via a secure channel.

$$RPW_i = h(PW_i \parallel K) \tag{3}$$

*2) Data Processing*
The $R$ computes $e_i$, $f_i$, $r_i$, $P_i$ and $R_i$ using (5), (4), (6) and (1) respectively.

$$f_i = h(ID_i \parallel R_i) \tag{4}$$
$$e_i = h(ID_i \parallel X_s) \oplus h(f_i \parallel RPW_i) \tag{5}$$
$$r_i = h(ID_i \parallel RPW_i) \tag{6}$$

*3) Card Preparation*
The $R$ stores $(e_i, f_i, r_i, P_i, h(.))$ on the $C_i$'s smart card and sends it to the $C_i$ via a secure channel.

*4) Finalization*
The $C_i$ enters $K$ into the smart card.

### B. Login Phase

During this phase, the user $C_i$ performs the following steps:

*1) Biometric Verification*
The $C_i$ inserts the smart card into a card reader and also provides $ID_i$, $PW_i$, $B_i$ to a specific device with fuzzy extractor and generates $R_i$ using (2). Then, the smart card computes $f_{ci}$ by placing provided $ID_i$ and calculated $R_i$ at (4). If $f_i = f_{ci}$, then the user $C_i$ passes biometric verification and continues the following steps. Otherwise, the session is terminated.

*2) Password Verification*
The smart card computes $RPW_i$ and $r_{ci}$ using (3) and by placing provided $ID_i$ and calculated $RPW_i$ at (6) respectively. It checks whether $r_i = r_{ci}$ or not. If they are equal, then $ID_i$ and $PW_i$ are verified and smart card performs the next step. Otherwise, the session is terminated.

*3) Login Request*
The smart card computes $M_1$, $M_2$ and $M_3$ using (7), (8) and (9) respectively.

$$M_1 = e_i \oplus h(f_i \parallel RPW_i) \tag{7}$$
$$M_2 = aP \text{ where } a \in Z_n^* \tag{8}$$

$$M_3 = h(M_1 \parallel M_2) \tag{9}$$

The $C_i$ sends a login request $\{ID_i, M_2, M_3\}$ to the server $S_i$.

### C. Authentication and Session Key Agreement Phase

During this phase, the user $C_i$ and the server $S_i$ perform the following steps:

*1) User ID Validation*
The $S_i$ checks the format of $ID_i$.

*2) Login Request Verification*
If $ID_i$ is valid, then the $S_i$ computes $M_4$ and $M_{c3}$ using following equations:

$$M_4 = h(ID_i \parallel X_s) \tag{10}$$
$$M_{c3} = h(M_4 \parallel M_2) \tag{11}$$

It checks whether $M_3 = M_{c3}$ or not. If they are equal, the $S_i$ accepts login request message and validity of the user $C_i$ is authenticated by the server $S_i$. Otherwise, the session is terminated.

*3) Mutual Authentication Request*
The server $S_i$ computes $M_5$ and $M_6$ using (12) and (13) respectively.

$$M_5 = bP \text{ where } b \in Z_n^* \tag{12}$$
$$M_6 = h(M_4 \parallel M_2 \parallel M_5) \tag{13}$$

It sends a mutual authentication message $\{M_5, M_6\}$ to the user $C_i$.

*4) Mutual Authentication Process*
After receiving reply, the user $C_i$ checks whether $M_6 = M_{c6}$ or not. The $M_{c6}$ is calculated as follow:

$$M_{c6} = h(M_1 \parallel M_2 \parallel M_5) \tag{14}$$

If they are equal, then the server $S_i$ is authenticated by the user $C_i$ and mutual authentication is completed.

*5) Session Key Generation*
The user $C_i$ and the server $S_i$ compute a shared key using (15).

$$SK = h(aM_5) = h(bM_2) = h(abP) \tag{15}$$

It is used for future confidential communication.

### D. Password Change Phase

During this phase, the user $C_i$ performs the following steps:

*1) Biometric Verification*
The $C_i$ inserts his smart card into a card reader and also provides $ID_i$, $PW_i$, $B_i$ to a specific device with fuzzy extractor and generates $R_i$ using (2). Then, smart card computes $f_{ci}$ by placing provided $ID_i$ and calculated $R_i$ at (4) and compares it with $f_i$ which is stored into the smart card. If $f_i = f_{ci}$, then the user $C_i$ passes biometric verification and continues the following steps.

*2) Password Verification*

The smart card computes $RPW_i$ and $r_{ci}$ using (3) and by placing provided $ID_i$ and calculated $RPW_i$ at (6) respectively and checks whether $r_i = r_{ci}$ or not. If they are equal, then $ID_i$ and $PW_i$ are verified and smart card performs the next step. The user inputs his new password $PW_{ni}$.

*3) Data Procession for New Password*

The smart card computes $RPW_{ni}$ and $r_{ni}$ using (3) and (6) respectively and by replacing $PW_i$ by $PW_{ni}$ and $RPW_i$ by $RPW_{ni}$. The $e_{ni}$ is calculated as follow:

$$e_{ni} = e_i \oplus h(f_i \parallel RPW_i) \oplus h(f_i \parallel RPW_{ni}) \quad (16)$$

*4) Finalization*

The smart card replaces $e_i$ and $r_i$ by $e_{ni}$ and $r_{ni}$ respectively to complete the phase.

IV. SECURITY ANALYSIS OF THE LI ET AL.'S SCHEME

In this section, we discuss security weaknesses of the scheme proposed by Li et al. [16]. We assume that the attacker $A_i$ can control the insecure channel.

A. *Password Guessing Attack Using Stolen Smart Card*

If $A_i$ can manage to steal the smart card, then he can extract information from the card by examining the power consumption signal [17]-[18]. He also can manage $ID_i$ by capturing one of the login request messages or simply using shoulder surfing technique. When the attacker manages to achieve information like $r_i$, $ID_i$, $K$, $h(.)$, he can conduct password guessing attack.

Consider that the $A_i$ selects a password $PW_{ai}$ from a massive database of passwords. Then, he computes $RPW_{ai}$ and $r_{ai}$ using (3) and (6) respectively and by replacing $PW_i$ with $PW_{ai}$ and $RPW_i$ with $RPW_{ai}$. He checks whether $r_i = r_{ai}$ or not. If they are equal, the selected password is correct. Otherwise, he repeats the process again.

B. *User impersonation Attack*

User impersonation attack can be launched after password guessing attack using stolen smart card. From previous discussion, we know that the attacker $A_i$ has ($e_i$, $f_i$, $r_i$, $P_i$, $h(.)$, $K$) from smart card and $ID_i$ as well as password $PW_{ai}$ from password guessing attack. Moreover, the parameters ($E_p(a,b)$, $n$, $P$) which are published by the registration center $R$, are stored in the smart card or published in a public domain. So, he can manage to gather these parameters, perform the following steps and try to login to the remote server $S_i$.

*1) Login Request by $A_i$*

The $A_i$ computes $RPW_{ai}$ using (3) and by replacing $PW_i$ with $PW_{ai}$. It also calculates $M_1$, $M_2$ and $M_3$ using (7), (8) and (9) respectively and by replacing $RPW_i$ with $RPW_{ai}$. It sends $\{ID_i, M_2, M_3\}$ to the server $S_i$.

*2) User ID Validation by $S_i$*

The $S_i$ receives message sent by the $A_i$ and verifies $ID_i$.

*3) Login Request Verification by $S_i$*

If $ID_i$ is valid, then the $S_i$ computes $M_4$ and $M_{c3}$ from (10) and (11) respectively. It checks whether $M_3 = M_{c3}$ or not. If they are equal, the server $S_i$ authenticates the $A_i$ as valid user.

*4) Mutual Authentication Request by $S_i$*

The server $S_i$ computes $M_5$ and $M_6$ using (12) and (13) respectively and sends a mutual authentication message $\{M_5, M_6\}$ to the attacker $A_i$.

*5) Mutual Authentication Process by $A_i$*

After receiving reply, the attacker $A_i$ calculates $M_{c6}$ using (14) and checks whether $M_6 = M_{c6}$ or not. If they are equal, then the server $S_i$ is authenticated by the attacker $A_i$ and the mutual authentication is completed.

*6) Session Key Generation by $A_i$ and $S_i$*

The attacker $A_i$ and the server $S_i$ compute shared key $SK$ using (15) and use it for future confidential communication.

C. *Security of the Secret Key*

The secret key $X_s$ remains stored in the server. Generally, the server stores this type of information in a database or in a file. Because of $X_s$ being unique for every user, the server $S_i$ has to maintain a mapping of $ID_i$ and $X_s$. According to the discussion in [6], information stored in the server database could be compromised. Therefore, this scheme is unable to provide the security of secret key.

D. *Server Masquerading Attack*

If the attacker $A_i$ can manage to steal the secret key $X_s$ as discussed in previous section, then it can launch attack as follow:

*1) Login Request by $C_i$*

The $C_i$ computes $RPW_i$, $M_1$, $M_2$ and $M_3$ using (3), (7), (8) and (9) and sends $\{ID_i, M_2, M_3\}$ to the $A_i$ (because the $A_i$ is masquerading as server).

*2) Login Request Verification by $A_i$*

The $A_i$ receives message sent by the $C_i$ and computes $M_4$ and $M_{c3}$ using (10) and (11) respectively and checks whether $M_3 = M_{c3}$ or not. If they are equal, then the $A_i$ authenticates the $C_i$ as valid user.

*3) Mutual Authentication Request by $A_i$*

The $A_i$ computes $M_5$ and $M_6$ using (12) and (13) respectively and sends a mutual authentication message $\{M_5, M_6\}$ to the user $C_i$.

*4) Mutual Authentication Process by $C_i$*

After receiving reply, the user $C_i$ calculates $M_{c6}$ using (14) and checks whether $M_6 = M_{c6}$ or not. If they are equal, then the $A_i$ (because the $C_i$ believes the $A_i$ as server) is authenticated by the user $C_i$ and the mutual authentication is completed.

*5) Session Key Generation by $C_i$ and $A_i$*

The user $C_i$ and attacker $A_i$ compute a shared key $SK$ using (15) and use it for future confidential communication.

E. *Password or Smart Card Recovery*

There is no password or smart card recovery phase in this scheme. According to the discussion of [5], the user $C_i$ can forget his password. If he forgets his password, there is no way he ever can get logged in. Moreover, if the attacker $A_i$ somehow can manage to steal the smart card of the user $C_i$, then the $C_i$ will not be able to recover the smart card.

*F. Mutual Authentication*

According to [15], if authentication scheme is insecure against user impersonation attack and server masquerading attack, then the authentication schemes cannot provide mutual authentication between the user and the remote server. Therefore, this scheme fails to provide mutual authentication according to the discussion in subsection IV(*B*) and IV(*D*).

## V. PROPOSED SCHEME

In this section, we first discuss some preliminary knowledge and then, we present our scheme.

*A. Background*

*1) Accessing Smart Card*

An attacker can access information of a smart card by using power analysis attack [17]-[18]. There are two types of power analysis attack: Simple Power Analysis (SPA) attack and Differential Power Analysis (DPA) attack. In SPA attack, the attacker visually inspects the system's power consumption. The DPA attack is more powerful than the SPA attack because the attacker does not need to know as many details about how the algorithm was implemented.

*2) Biometric Key Generation*

Biometric cryptosystem and cancelable biometrics represent emerging technologies to release biometric keys as well as provide privacy to biometric templates [19]. Dodis et al. claimed that strong biometric keys can be generated from biometric templates using fuzzy extractor [20]. An efficient cancelable biometric key generation scheme for cryptographic use that can generate 256 bit keys from fingerprint templates is proposed in [21] and an efficient approach for non-invertible cryptographic key generation from cancelable fingerprint biometrics is proposed at [22]. Jagadeesan et al. [23] proposed a cryptographic key generation scheme from multiple biometric modalities that can generate 256 bit key. One of these efficient key generation methods can be used in our proposed scheme.

*3) Hash Function*

A cryptographic hash function is a mathematical algorithm that maps data of arbitrary size to a bit string of a fixed size (a hash function) which is designed to be one-way function, that is, a function which is infeasible to invert [24]. The example of hash functions are MD5, SHA-1, SHA-2 family, SHA-3 family, etc. The MD5 and SHA-1 are no longer recommended due to security reason. Either SHA-2 family or SHA-3 family should be used to generate message digest.

*4) Advanced Encryption Standard (AES)*

AES is a symmetric key algorithm where a single key named private key is used for both encryption and decryption purpose [25]. It is also known as Rijndael. It is based on a design principle known as substitution-permutation network. It is a block cipher with three different key lengths: 128, 192 and 256 bits. The degree of security relies on the key length. Despite the different key length AES operate on 128 bits plain text block and generates 128 bits cipher text. The performance of AES is very convincing. It is a high speed algorithm and also requires low RAM. Therefore, it can be implemented from 8 bit smart card to high performance computer.

*B. Our Proposed Scheme*

Our proposed scheme consists of six phases that includes server registration phase, user registration phase, login and authentication phase, password change phase, password recovery phase, and smart card recovery phase. Also, we have few assumptions under which our proposed scheme worked properly. They are given bellow:

- User or server id is unique but not a secret.
- Password is secret but it may not be unique.
- Biometric key is unique and very hard to be copied, shared and distributed.
- AES keys are secret and the attacker cannot steal them.
- Smart card can be stolen and information stored in smart card can be revealed.
- The attacker cannot manage to steal smart card, password and biometric key at the same time.
- The attacker cannot take over secure channel.
- The attacker has control over insecure channel.

*C. Server Registration Phase*

During the server registration phase, the server $S_i$ and the registration center $R$ will perform the following steps:

*1) Registration Request by $S_i$*

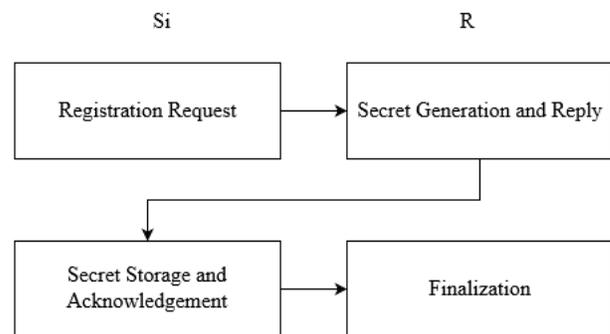

Fig. 1. Server registration phase which involves the server $S_i$ and the registration center $R$

The server sets $Stat = Register$ and sends $\{SID_i, Stat\}$ to the registration center $R$.

*2) Secret Generation and Reply by R*

The $R$ receives $\{SID_i, Stat\}$ from the $S_i$. If $Stat = Register$, then it checks into its database. If the server $S_i$ is already registered, then it discards the process. Otherwise, it chooses a secret random string $R_{n1}$ and generates $K_s$ using (17).

$$K_s = h(SID_i \parallel R_{n1}) \qquad (17)$$

It sets $Stat = Accept$ and sends $\{SID_i, K_s, Stat\}$ to the $S_i$ through a secure channel.

3) *Secret Storage and Acknowledgement by $S_i$*
The $S_i$ receives $\{SID_i, K_s, Stat\}$ from the $R$ and checks $SID_i$. If $SID_i$ matches with its own and $Stat = Accept$, then it calculates $EK_s$ using (18).

$$EK_s = E_{aes}(K_s, SK_{aes}) \qquad (18)$$

It stores $EK_s$ into its database. It sets $Stat = Ack$ and sends $\{SID_i, Stat\}$ to the $R$.

4) *Finalization by R*
After receiving acknowledgement $\{SID_i, Stat\}$ from the $S_i$, the $R$ calculates $HK_s$ as follow:

$$HK_s = E_{aes}(K_s, RK_{aes}) \qquad (19)$$

It stores $\{SID_i, HK_s\}$ into its database. The server registration phase is illustrated in Fig. 1.

D. *User Registration Phase*

During user registration phase, the user $C_i$, the registration center $R$, and the server $S_i$ will perform the following steps:

1) *Registration Request by $C_i$*
The user $C_i$ needs to choose his user identification $ID_i$, password $PW_i$, recovery contact $R_{cont}$, collect server identification $SID_i$ which is published publicly, and imprint his biometrics in a specific device which can generate biometric key $B_i$ form biometrics. He also calculates $BP_i$ as follow:

$$BP_i = h(PW_i \parallel B_i) \qquad (20)$$

Then, he sets $Stat = Register$ and sends $\{ID_i, BP_i, SID_i, R_{cont}, Stat\}$ to the registration center $R$ through a secure channel.

2) *Registration Request by R*
The registration center $R$ receives message $\{ID_i, BP_i, SID_i, R_{cont}, Stat\}$ from the $C_i$. If $Stat = Register$ and $SID_i$ is already registered, then it generates a secret random string $W$ and calculates $TX_s$ using (21).

$$TX_s = W \parallel BP_i \qquad (21)$$

The $R$ sends $\{ID_i, SID_i, TX_s, Stat\}$ to the server $S_i$ through a secure channel.

3) *Secret Generation, Storage and Reply by $S_i$*
The server $S_i$ receives $\{ID_i, SID_i, TX_s, Stat\}$ from the $R$. It verifies $SID_i$. If the verification passes, then it proceeds; otherwise, it discards the request. If the verification passes and $Stat = Register$, then it calculates $K_s$, $X_s$ and $SX_i$ using (22), (23) and (24) respectively.

$$K_s = D_{aes}(EK_s, SK_{aes}) \qquad (22)$$
$$X_s = h(K_s \parallel TX_s) \qquad (23)$$

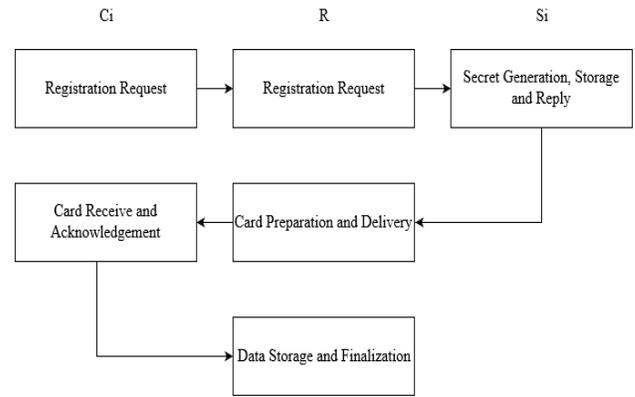

Fig. 2. User registration phase which involves the user $C_i$, the registration center $R$ and the server $S_i$

$$SX_i = E_{aes}(X_s, SK_{aes}) \qquad (24)$$

Then, it stores $\{ID_i, SX_i\}$ into its database, sets $Stat = Complete$ and sends $\{ID_i, SID_i, Stat\}$ to the registration center $R$ through a secure channel.

4) *Card Preparation and Delivery by R*
The $R$ receives $\{ID_i, SID_i, Stat\}$ from the $S_i$. It verifies $ID_i$ and $SID_i$. If verification passes and $Stat = Complete$, then it confirms that the registration process in the server has completed. Then, it calculates $K_s$ and $TC_s$ using (25) and (26) respectively.

$$K_s = D_{aes}(HK_s, RK_{aes}) \qquad (25)$$
$$TC_s = K_s \parallel W \qquad (26)$$

It stores $\{ID_i, SID_i, BP_i, TC_s\}$ into a smart card and distributes it to the $C_i$ through a secure channel.

5) *Card Receive and Acknowledgement by $C_i$*
The $C_i$ receives the smart card from the $R$. Then, he puts the smart card into a card reader and checks $ID_i$ and $SID_i$. If they are correct, then he provides $PW_i$ and imprints his biometrics to the specific device to generate biometric key $B_i$. Then, he calculates $BP_{ci}$ by putting $PW_i$ and $B_i$ at (20) and compares whether $BP_i = BP_{ci}$ or not. If the verification passes, then he accepts the card, sets $Stat = Accept$ and sends $\{ID_i, SID_i, Stat\}$ to the $R$ through a secure channel.

$$QX_i = E_{aes}(TC_s, B_i) \qquad (27)$$

He calculates $QX_i$ using (27) and replaces $TC_s$ in the smart card. He also removes $BP_i$ from the card. If the verification fails, he rejects the card, sets $Stat = Reject$ and sends a failure message $\{ID_i, SID_i, Stat\}$ to the $R$ and discards process.

6) *Data Storage and Finalization by R*
The $R$ receives $\{ID_i, SID_i, Stat\}$ from the $C_i$. It verifies $ID_i$ and $SID_i$. If verification passes, then it checks $Stat$. If $Stat = Accept$, then it confirms that the card has reached to its designated user. Then, it calculates $R_{cov}$, $EX_i$ and $UX_i$ as follow:

$$R_{cov} = E_{aes}(R_{cont}, RK_{aes}) \quad (28)$$
$$EX_i = E_{aes}(TX_s, RK_{aes}) \quad (29)$$
$$UX_i = E_{aes}(TC_s, RK_{aes}) \quad (30)$$

It stores $\{ID_i, SID_i, UX_i, EX_i, R_{cov}\}$ into its database. If $Stat = Reject$ or if the verification fails, then it discards the process and sets $Stat = Deregister$ and sends the $S_i$ a message $\{ID_i, SID_i, Stat\}$ through a secure channel to deregister the user. When $S_i$ receives such message, then it deletes the corresponding data from its database. The user registration phase is illustrated in Fig. 2.

E. *Login and Authentication Phase*

During this phase, the user $C_i$ and the server $S_i$ will perform the following steps:

1) *Login Request by $C_i$*
The user $C_i$ inserts his smart card into the card reader. He also provides his $ID_i$, $PW_i$ and imprints his biometrics to a specific device to generate biometric key $B_i$. Then, the user $C_i$ verifies $ID_i$. If the verification fails, then he terminates the session. After that, he calculates $BP_i$, $TC_s$ and $X_s$ using (10), (31) and (32) respectively.

$$TC_s = D_{aes}(QX_i, B_i) \quad (31)$$
$$X_s = h(TC_s \parallel BP_i) \quad (32)$$
$$M_1 = h(X_s \parallel R_{n2}) \quad (33)$$
$$M_2 = h(ID_i \parallel X_s) \oplus R_{n2} \quad (34)$$

He generates a secret random string $R_{n2}$ and calculates $M_1$ and $M_2$ using (33) and (34) respectively. Then, he sets $Stat = Login$ and sends $\{ID_i, SID_i, M_1, M_2, Stat\}$ to the server $S_i$.

2) *Verification and Mutual Authentication Request by $S_i$*
The server $S_i$ receives login message $\{ID_i, SID_i, M_1, M_2, Stat\}$ from the user $C_i$. It verifies $ID_i$ of the message with stored $ID_i$ and $SID_i$ with its server id. If the verification fails, then it terminates the session. If the verification passes and $Stat = Login$, then it proceeds. It calculates $X_s$ and $R_{n2}$ using (35) and (36) respectively and $M_3$ by placing calculated $X_s$ and $R_{n2}$ at (33).

$$X_s = D_{aes}(SX_i, SK_{aes}) \quad (35)$$
$$R_{n2} = M_2 \oplus h(ID_i \parallel X_s) \quad (36)$$
$$M_5 = h(ID_i \parallel X_s \parallel R_{n2}) \oplus R_{n3} \quad (37)$$

Then, it compares whether $M_1 = M_3$ or not. If they are not equal, it terminates the session. Otherwise, the user is authenticated. It generates a secret random string $R_{n3}$. It calculates $M_4$ by replacing $R_{n2}$ with $R_{n3}$ at (33) and $M_5$ using (37). Then, it sets $Stat = Auth$ and sends $\{ID_i, SID_i, M_4, M_5, Stat\}$ to the user $C_i$.

3) *Mutual Authentication and Acknowledgement by $C_i$*
The user $C_i$ receives message $\{ID_i, SID_i, M_4, M_5, Stat\}$ from the $S_i$. He verifies $ID_i$ and $SID_i$ of the message. If the verification fails, then he terminates the session. If the verification passes and $Stat = Auth$, then he proceeds. He calculates $R_{n3}$ using (38) and $M_6$ by putting $X_s$ and calculated $R_{n3}$ at (33).

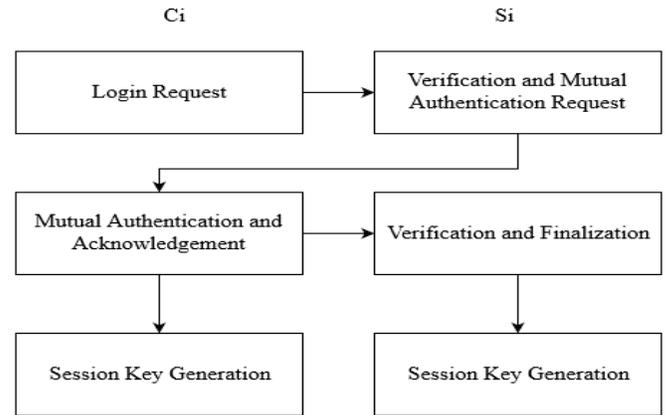

Fig. 3. Login and authentication phase which involves the user $C_i$ and the server $S_i$

$$R_{n3} = M_5 \oplus h(ID_i \parallel X_s \parallel R_{n2}) \quad (38)$$
$$M_7 = h(X_s \parallel R_{n2} \parallel R_{n3}) \quad (39)$$

He compares whether $M_4 = M_6$ or not. If they are not equal, then he terminates the session. Otherwise, the server is authenticated. The user calculates $M_7$ using (39) and sets $Stat = Auth$ and sends $\{ID_i, SID_i, M_7, Stat\}$ to the server $S_i$. If the session is terminated, he sends login request again.

4) *Verification and Finalization by $S_i$*
The server $S_i$ receives $\{ID_i, SID_i, M_7, Stat\}$ from the $C_i$. Then, it checks $ID_i$ and $SID_i$. If $ID_i$ is desired user id, $SID_i$ is desired server id and $Stat = Auth$, then it calculates $M_8$ by putting $X_s$, $R_{n2}$ and $R_{n3}$ at (39). It compares whether $M_7 = M_8$ or not. If they are not equal, it discards the message and terminates the session. Otherwise, the authentication is completed.

5) *Session Key Generation by $C_i$ and $S_i$*
The $S_i$ and the $C_i$ both calculate the session key for further secret communication. The session key is calculated as follow:

$$K_{ses} = h(R_{n2} \parallel R_{n3}) \quad (40)$$

The login and authentication phase is illustrated in Fig. 3.

F. *Password Change Phase*

During password change phase, the user $C_i$, the registration center $R$, and the server $S_i$ will perform the following steps:

1) *Password Change Request by $C_i$*
The user $C_i$ inserts his smart card into the card reader. He also provides his $ID_i$, $PW_i$ and imprints his biometrics into a specific device to generate biometric key $B_i$. Then, the user $C_i$ verifies $ID_i$. If the verification fails, he discards the process. Otherwise, he calculates $BP_i$, $TC_s$, $X_s$ and $TCX_s$ using (20), (31), (32) and (41) respectively.

$$TCX_s = TC_s \oplus X_s \quad (41)$$

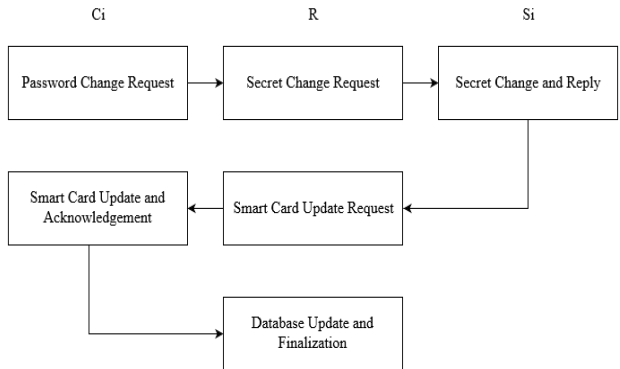

Fig. 4. Password change phase which involves the user $C_i$, the registration center $R$ and the server $S_i$

He also provides a new password $PW_{ni}$. Then, he calculates $BP_{ni}$ by replacing $PW_i$ with $PW_{ni}$ at (20), sets $Stat = Passchange$ and sends $\{ID_i, TCX_s, BP_{ni}, SID_i, Stat\}$ to the $R$ through a secure channel.

2) *Secret Change Request by R*

The $R$ receives message $\{ID_i, TCX_s, BP_{ni}, SID_i, Stat\}$ from the $C_i$. It verifies $ID_i$ and $SID_i$ with its database. If verification fails, then it discards the request. If the verification passes and $Stat = Passchange$, then it calculates $TC_s, K_s, TX_s$ and $X_s$ using (42), (25), (43) and (44) respectively and $X_{cs}$ by putting calculated $K_s$ and $TX_s$ at (23).

$$TC_s = D_{aes}(UX_i, RK_{aes}) \qquad (42)$$
$$TX_s = D_{aes}(EX_i, RK_{aes}) \qquad (43)$$
$$X_s = TCX_s \oplus TC_s \qquad (44)$$

It compares whether $X_{cs} = X_s$ or not. If they are not equal, then the request is discarded. If they match, then it sets $Stat = Passchange$, chooses a secret random string $R_{n4}$, calculates $TX_{ns}$ by replacing $W$ and $BP_i$ with $R_{n4}$ and $BP_{ni}$ respectively at (21) and sends $\{ID_i, SID_i, TX_s, TX_{ns}, Stat\}$ to the server $S_i$ through a secure channel. If the process is discarded, then the $R$ sets $Stat = Fail$ and sends failure message $\{ID_i, Stat\}$ to the user $C_i$. If the $C_i$ receives this failure message, then he sends password change request again.

3) *Secret Change and Reply by $S_i$*

The server $S_i$ receives $\{ID_i, SID_i, TX_s, TX_{ns}, Stat\}$ from the $R$. It verifies $ID_i$ and $SID_i$. If the verification passes, then it proceeds. Otherwise, it discards the request. It calculates $K_s$ and $X_s$ using (22) and (35) respectively, and $X_{cs}$ by putting calculated $K_s$ and $TX_s$ at (23), and compares whether $X_s = X_{cs}$ or not. If they are not equal, then it discards the request. If they match, then it calculates $X_{ns}$ and $SX_{ni}$ by replacing $TX_s$ with $TX_{ns}$ and $X_s$ with $X_{ns}$ at (23) and (24) respectively. It replaces $SX_i$ with $SX_{ni}$ in its database. Then, it sets $Stat = Complete$ and sends $\{ID_i, SID_i, Stat\}$ to the $R$ through a secure channel. If the process is discarded, then it sets $Stat = Fail$ and sends a failure message $\{ID_i, SID_i, Stat\}$ to the $R$. If the $R$ receives this failure message, then it sends secret change request again.

4) *Smart Card Update Request by R*

The $R$ receives $\{ID_i, SID_i, Stat\}$ from the $S_i$. It verifies $ID_i$ and $SID_i$. If verification passes and $Stat = Complete$, then it assumes that the update process in the server has completed. Otherwise, if the verification fails, then it discards the process. When it gets the confirmation of the completion of the server update, then it calculates $TC_{ns}$ by replacing $W$ with $R_{n4}$ at (26), and $TC_s$ using (42), sets $Stat = Complete$ and sends $\{ID_i, SID_i, TC_{ns}, TC_s, Stat\}$ to the user $C_i$ through a secure channel. If the process is discarded, it sets $Stat = Fail$ and sends a failure message $\{ID_i, SID_i, Stat\}$ to the $S_i$. If the $S_i$ receives this failure message, then it simply reverts database changes. The $R$ sends secret change request to the $S_i$ again.

5) *Smart Card Update and Acknowledgement by $C_i$*

The user $C_i$ receives message $\{ID_i, SID_i, TC_{ns}, TC_s, Stat\}$ from the $R$. Then he checks $ID_i$ and $SID_i$. If they are correct and $Stat = Complete$, then he calculates $TC_{cs}$ using (31) and compares whether $TC_s = TC_{cs}$ or not. If the verification passes, then he sets $Stat = Complete$ and sends $\{ID_i, SID_i, Stat\}$ to the $R$ through a secure channel. Then, he calculates $QX_{ni}$ by replacing $TC_s$ with $TC_{ns}$ at (27) and replaces $QX_i$ with $QX_{ni}$ in the smart card. If any of the verifications fails, then he rejects the reply. If the reply is rejected, then he discards the process, sets $Stat = Fail$ and sends a failure message $\{ID_i, SID_i, Stat\}$ to the $R$ and retries password change again. If the $R$ receives this failure message, then it sends a failure message to the $S_i$ to revert database changes and the $S_i$ does accordingly.

6) *Database Update and Finalization by R*

The $R$ receives $\{ID_i, SID_i, Stat\}$ from the user $C_i$ and verifies $ID_i$ and $SID_i$. If verification passes and $Stat = Complete$, then it confirms that the card has updated and it calculates $UX_{ni}$ and $EX_{ni}$ by replacing $TC_s$ with $TC_{ns}$ and $TX_s$ with $TX_{ns}$ at (30) and (29) respectively. It also replaces $UX_i$ with $UX_{ni}$ and $EX_i$ with $EX_{ni}$ into its own database. Otherwise, if the verification fails, then it discards the process. The $R$ sets $Stat = Fail$ and sends a failure message $\{ID_i, SID_i, Stat\}$ to the $S_i$ and the $C_i$ to revert their changes and they act accordingly. The password change phase is illustrated in Fig. 4.

G. *Password Recovery Phase*

If the user $C_i$ forgets his password, then $C_i$, the registration center $R$, and the server $S_i$ will have to perform the following steps:

1) *Password Recovery Request by $C_i$*

The user $C_i$ needs to provide his user identification $ID_i$ and collect server identification $SID_i$ which is published publicly. Then he sets $Stat = Recovery$ and sends message $\{ID_i, SID_i, Stat\}$ to the $R$ through a secure channel.

2) *User Verification Request by R*

The $R$ receives message $\{ID_i, SID_i, Stat\}$ from the user $C_i$. It verifies $ID_i$ and $SID_i$. If verification passes and $Stat = Recovery$, then it proceeds. Otherwise, if verification process fails, then it terminates the process. It generates a secret random string $R_{n5}$ and calculates $R_{cont}$ as follow:

$$R_{cont} = D_{aes}(R_{cov}, RK_{aes}) \qquad (45)$$

It sets $Stat = Verify$ and sends a message $\{ID_i, SID_i, R_{n5}, Stat\}$ to the recovery contact ($R_{cont}$) of the $C_i$ through a secure channel.

3) *Verification Reply by $C_i$*

The user $C_i$ receives message $\{ID_i, SID_i, R_{n5}, Stat\}$ from the $R$. Then he checks $ID_i$ and $SID_i$. If they are correct and $Stat = Verify$, then he proceeds. He sets $Stat = Verify$, chooses a new password $PW_{ni}$, calculates $BP_{ni}$ by replacing $PW_i$ with $PW_{ni}$ at (20) and sends $\{ID_i, SID_i, R_{n5}, BP_{ni}, Stat\}$ to the $R$ through a secure channel. Otherwise, if verification fails, then he simply discards the message and sends recovery request again.

4) *Finalization of Verification and Secret Change Request by Registration Center R*

The $R$ receives message $\{ID_i, SID_i, R_{n5}, BP_{ni}, Stat\}$ from the user $C_i$. It verifies $ID_i$ and $SID_i$. If verification passes and $Stat = Verify$, then it verifies $R_{n5}$. If it holds, then it confirms that the user $C_i$ is valid. Otherwise, if any of the verifications fails, then it discards the message. If the user $C_i$ is valid, then it chooses a secret random string $R_{n6}$, calculates $TX_s$ using (43), and $TX_{ns}$ by replacing $W$ and $BP_i$ with $R_{n6}$ and $BP_{ni}$ respectively at (21). Then, it sets $Stat = Recovery$ and sends $\{ID_i, SID_i, TX_s, TX_{ns}, Stat\}$ to the server $S_i$ through a secure channel.

5) *Secret Change and Reply by $S_i$*

The server $S_i$ receives $\{ID_i, SID_i, TX_s, TX_{ns}, Stat\}$ from the $R$. It verifies $ID_i$ and $SID_i$. If verification passes, then it proceeds; otherwise, it discards the request. It calculates $K_s$ and $X_s$ using (22) and (35) respectively, and $X_{cs}$ by putting calculated $K_s$ and $TX_s$ at (23), and compares whether $X_s = X_{cs}$ or not. If they are not equal, then it discards the request. If they match, then it calculates $X_{ns}$ and $SX_{ni}$ by replacing $TX_s$ with $TX_{ns}$ and $X_s$ with $X_{ns}$ at (23) and (24) respectively and replaces $SX_i$ with $SX_{ni}$ in its database. Then, it sets $Stat = Done$ and sends $\{ID_i, SID_i, Stat\}$ to the $R$ through a secure channel. If the process is discarded, then it sets $Stat = Fail$ and sends a failure message $\{ID_i, SID_i, Stat\}$ to the $R$. If the $R$ receives the failure message, then it sends secret change request again.

6) *Smart Card Update Request by R*

The $R$ receives $\{ID_i, SID_i, Stat\}$ from the $S_i$. It verifies $ID_i$ and $SID_i$. If verification passes and $Stat = Done$, then it assumes that the update process in the server has completed. Otherwise, if verification fails, then it discards the process. When it has the confirmation of the completion of the server update, then it calculates $TC_{ns}$ by replacing $W$ with $R_{n6}$ at (26), and $TC_s$ using (42). It sets $Stat = Done$ and sends $\{ID_i, SID_i, TC_{ns}, TC_s, Stat\}$ to the user $C_i$ through a secure channel. If the process is discarded, it sets $Stat = Done$ and sends a failure message $\{ID_i, SID_i, Stat\}$ to the $S_i$. If the $S_i$ receives this failure message, then it simply reverts database changes. The $R$ sends secret change request to the $S_i$ again.

7) *Smart Card Update and Acknowledgement by $C_i$*

The user $C_i$ receives message $\{ID_i, SID_i, TC_{ns}, TC_s, Stat\}$ from the $R$. Then he checks $ID_i$ and $SID_i$. If they are correct and $Stat = Done$, then he calculates $TC_{cs}$ using (31) and compares whether $TC_s = TC_{cs}$ or not. If verification passes, then he sets $Stat = Complete$ and sends $\{ID_i, SID_i, Stat\}$ to the $R$ through a secure channel. Then, he calculates $QX_{ni}$ by replacing $TC_s$ with $TC_{ns}$ at (27) and replaces $QX_i$ with $QX_{ni}$ in the smart card. If any of the verifications fails, then he rejects the reply. If the reply is rejected, he discards the process, sets $Stat = Fail$ and sends a failure message $\{ID_i, SID_i, Stat\}$ to the $R$ and retries password recovery again. If the $R$ receives this failure message, then it sends a failure message to the $S_i$ to revert database changes and the $S_i$ does accordingly.

8) *Database Update and Finalization by R*

The $R$ receives $\{ID_i, SID_i, Stat\}$ from the user $C_i$. It verifies $ID_i$ and $SID_i$. If verification passes and $Stat = Complete$, then it confirms that the card has updated and it calculates $UX_{ni}$ and $EX_{ni}$ by replacing $TC_s$ with $TC_{ns}$ and $TX_s$ with $TX_{ns}$ at (30) and (29) respectively. It also replaces $UX_i$ with $UX_{ni}$ and $EX_i$ with $EX_{ni}$ into its own database. Otherwise, if verification fails, then it discards the process. The $R$ sets $Stat = Fail$ and sends a failure message $\{ID_i, SID_i, Stat\}$ to the $S_i$ and the $C_i$ to revert their changes and they act accordingly. The Password recovery phase is illustrated in Fig. 5.

*H. Smart Card Recovery Phase*

If the user $C_i$ loses his smart card, then he, the registration center $R$ and the server $S_i$ will have to perform the following steps:

1) *Smart Card Recovery Request by $C_i$*

The user $C_i$ needs to provide his user identification $ID_i$ and collect server identification $SID_i$ which is published publicly. Then, he sets $Stat = RecoveryS$ and sends message $\{ID_i, SID_i, Stat\}$ to the $R$ through a secure channel.

2) *User Verification Request by R*

The $R$ receives message $\{ID_i, SID_i, Stat\}$ from the user $C_i$. It verifies $ID_i$ and $SID_i$. If verification passes and $Stat = RecoveryS$, then it proceeds. Otherwise, if verification process fails, then it terminates the process. It generates a secret random string $R_{n7}$ and calculates $R_{cont}$ using (45). It sets $Stat = VerifyS$ and sends a message $\{ID_i, SID_i, R_{n7}, Stat\}$ to the recovery contact ($R_{cont}$) of the $C_i$ through a secure channel.

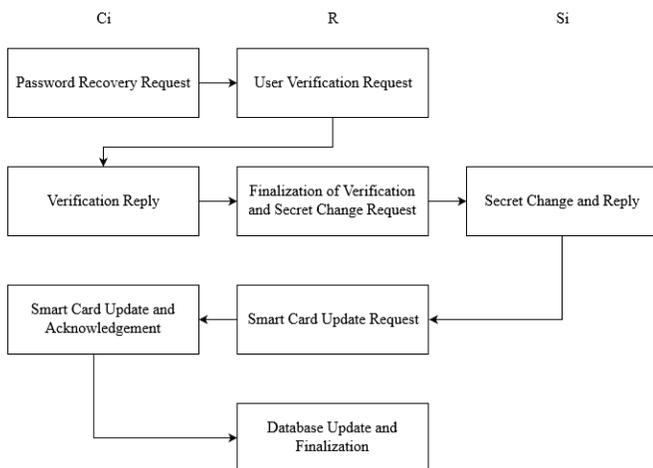

Fig. 5. Password recovery phase which involves the user $C_i$, the registration center $R$ and the server $S_i$

3) *Verification Reply by $C_i$*

The user $C_i$ receives message $\{ID_i, SID_i, R_{n7}, Stat\}$ from the $R$. Then, he checks $ID_i$ and $SID_i$. If they are correct and $Stat = VerifyS$, then he proceeds. He sets $Stat = VerifyS$, chooses a new password $PW_{ni}$, calculates $BP_{ni}$ by replacing $PW_i$ with $PW_{ni}$ at (20) and sends $\{ID_i, SID_i, R_{n7}, BP_{ni}, Stat\}$ to the $R$ through a secure channel. Otherwise, if verification fails, then he simply discards the message and sends recovery request again.

4) *Finalization of Verification and Secret Change Request by Registration Center R*

The $R$ receives message $\{ID_i, SID_i, R_{n7}, BP_{ni}, Stat\}$ from the user $C_i$. It verifies $ID_i$ and $SID_i$. If verification passes and $Stat = VerifyS$, then it verifies $R_{n7}$. If it holds, then it confirms that the user $C_i$ is valid. Otherwise, if any of the verifications fails, it discards the message. If the user $C_i$ is valid, then it chooses a secret random string $R_{n8}$, calculates $TX_s$ using (43) and $TX_{ns}$ by replacing $W$ and $BP_i$ with $R_{n8}$ and $BP_{ni}$ respectively at (21). Then, it sets $Stat = RecoveryS$ and sends $\{ID_i, SID_i, TX_s, TX_{ns}, Stat\}$ to the server $S_i$ through a secure channel.

5) *Secret Change and Reply by $S_i$*

The server $S_i$ receives $\{ID_i, SID_i, TX_s, TX_{ns}, Stat\}$ from the $R$. It verifies $ID_i$ and $SID_i$. If verification passes and $Stat = RecoveryS$, then it proceeds; otherwise, it discards the request. It calculates $K_s$ and $X_s$ using (22) and (35) respectively, and $X_{cs}$ by putting calculated $K_s$ and $TX_s$ at (23), and compares whether $X_s = X_{cs}$ or not. If they are not equal, then it discards the request. If they match, then it calculates $X_{ns}$ and $SX_{ni}$ by replacing $TX_s$ with $TX_{ns}$ and $X_s$ with $X_{ns}$ at (23) and (24) respectively, and replaces $SX_i$ with $SX_{ni}$ into its database. Then, it sets $Stat = DoneS$ and sends $\{ID_i, SID_i, Stat\}$ to the $R$ through a secure channel. If the process is discarded, then it sets $Stat = Fail$ and sends a failure message $\{ID_i, SID_i, Stat\}$ to the $R$. If the $R$ receives this failure message, then it sends secret change request again.

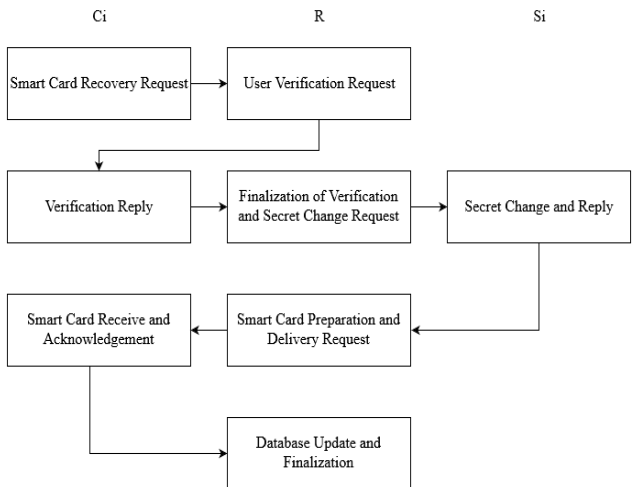

Fig. 6. Smart card recovery phase which involves the user $C_i$, the registration center $R$ and the server $S_i$

6) *Smart Card Preparation and Delivery by R*

The $R$ receives $\{ID_i, SID_i, Stat\}$ from the $S_i$ and verifies $ID_i$ and $SID_i$. If verification passes and $Stat = DoneS$, then it assumes that the update process in the server has completed. Otherwise, if verification fails, then it discards the process. When it has the confirmation of the completion of the server update, it calculates $TC_{ns}$ by replacing $W$ with $R_{n8}$ at (26) and $TC_s$ using (42). It stores $\{ID_i, SID_i, BP_{ni}, TC_{ns}\}$ into a smart card and distributes it to the user $C_i$ through a secure channel. If the process is discarded, it sets $Stat = Fail$ and sends a failure message $\{ID_i, SID_i, Stat\}$ to the $S_i$. When the $S_i$ receives this failure message, it simply reverts database changes. The $R$ sends secret change request to the $S_i$ again.

7) *Smart Card Receive and Acknowledgement by $C_i$*

The $C_i$ receives smart card from the $R$. Then, the $C_i$ puts his smart card into a card reader. Then, he checks $ID_i$ and $SID_i$. If they are correct, then the $C_i$ provides $PW_{ni}$ and imprints his biometrics to the specific device to generate biometric key $B_i$. Then, he calculates $BP_{ci}$ by replacing $PW_i$ with $PW_{ni}$ at (20) and compares whether $BP_{ni} = BP_{ci}$ or not. If verification passes, then he accepts the card, sets $Stat = AcceptS$ and sends $\{ID_i, SID_i, Stat\}$ to the $R$ through a secure channel. Then, he calculates $QX_{ni}$ by replacing $TC_s$ with $TC_{ns}$ at (27), and replaces $TC_{ns}$ with $QX_{ni}$ in the smart card. He also removes $BP_{ni}$ from the card. If verification fails, he discards the process, sets $Stat = Fail$, sends a failure message $\{ID_i, SID_i, Stat\}$ to the $R$ trough a secure channel and sends recovery request again. If the $R$ receives this failure message, then it sends a failure message to the $S_i$ to revert database changes and the $S_i$ does it accordingly.

8) *Database Update and Finalization by R*

The $R$ receives $\{ID_i, SID_i, Stat\}$ from the user $C_i$. It verifies $ID_i$ and $SID_i$. If verification passes and $Stat = AcceptS$, then it confirms that the card has reached to its designated user and it calculates $UX_{ni}$ and $EX_{ni}$ by replacing $TC_s$ with $TC_{ns}$ and $TX_s$ with $TX_{ns}$ at (30) and (29) respectively. It also replaces $UX_i$ with $UX_{ni}$ and $EX_i$ with $EX_{ni}$ into its own database. Otherwise, if verification fails, then it discards the process. The $R$ sets $Stat = Fail$ and sends a failure message $\{ID_i, SID_i, Stat\}$ to the $S_i$ and the $C_i$ to revert their changes and they act accordingly. The Smart card recovery phase is illustrated in Fig. 6.

VI. SECURITY ANALYSIS OF THE PROPOSED SCHEME

In this section, we will show how our scheme can resist security attacks and prevent the attacker $A_i$ to cause any potential harm.

A. *Password Guessing Attack*

If the attacker $A_i$ can manage to steal the smart card, then he can manage to extract information from the card by examining the power consumption signal [17]-[18]. He can collect information like $QX_i$ from it. The $QX_i$ is an encrypted information and contains $TC_s$ within it. Moreover, we can clearly see from (26) that $TC_s$ holds no information about password. Also, he can manage to get messages like $\{ID_i,$

$SID_i, M_1, M_2, Stat\}, \{ID_i, SID_i, M_4, M_5, Stat\}, \{ID_i, SID_i, M_7, Stat\}$ and $\{ID_i, SID_i, Stat\}$ from login and authentication phase by eavesdropping the insecure channel. The $M_1, M_2, M_5$ and $M_7$ are calculated using (33), (34), (37) and (39) respectively. The $M_4$ is calculated by replacing $R_{n2}$ with $R_{n3}$ at (33). But, none of these messages contain password directly. So, there is no way the attacker can conduct password guessing attack by trial and error basis.

*B. Secret Key Stealing*

The master secret key $X_s$ is not stored in the smart card. To generate $X_s$, the attacker needs to decrypt $QX_i$ to get $TC_s$ and also needs biometric key $B_i$ and password $PW_i$. From section A, we can say that the attacker cannot manage to get password $PW_i$. At the same time biometric key $B_i$ is very hard to copy. Therefore, the attacker cannot generate $X_s$ from smart card.

At the trusted registration center, we store $EX_i$, $UX_i$ and $HK_s$. From (29), (30) and (19), we can see all these data are encrypted. Unless the attacker can manage to get $RK_{aes}$, it is almost impossible for him to get $K_s$, $TC_s$ and $TX_s$. Without these, he cannot generate master secret key $X_s$.

At the server, we store $SX_i$ and $EK_s$. From (24) and (18), we can see that all these data are encrypted. Unless the attacker can manage to get $SK_{aes}$, it is almost impossible for him to get $X_s$.

Because $RK_{aes}$ and $SK_{aes}$ both are AES keys and kept secret by the trusted registration center and the server respectively, we assume that the attacker won't be able to get them. Therefore, the attacker cannot get $X_s$ from the trusted registration center and the server.

The attacker $A_i$ can gather login messages like $M_1, M_2, M_4, M_5$ and $M_7$ by forging into the insecure channel which is used during login and authentication phase. Here, $R_{n2}$ and $R_{n3}$ are two secret random strings generated during login, and authentication phase and they change every time during message generation. Therefore, there is no way to collect these strings. This is why, it is nearly impossible for an attacker to guess $X_s$ from these messages.

*C. User Impersonation Attack*

To conduct user impersonation attack, the attacker $A_i$ needs to send login request $\{ID_i, SID_i, M_1, M_2, Stat\}$. More precisely, he needs to generate $M_1$ and $M_2$ using (33) and (34) respectively. From previous discussion at subsection VI(*B*), we know that the attacker cannot manage $X_s$ and $R_{n2}$. So, it is not possible for him to generate $M_1$ and $M_2$. Therefore, he cannot impersonate as user.

*D. Server Masquerading Attack*

To conduct server masquerading attack, the attacker $A_i$ need to send mutual authentication request $\{ID_i, SID_i, M_4, M_5, Stat\}$. More precisely, he needs to generate $M_4$ and $M_5$ using (33) and (37) respectively. From previous discussion at subsection VI(*B*), we already know that the attacker cannot manage $X_s$, $R_{n2}$ and $R_{n3}$. So, it is not possible for him to generate $M_4$ and $M_5$. Therefore, he cannot masquerade as server.

*E. Replay Attack*

In this case, the attacker $A_i$ may use captured $M_1$ and $M_2$ for sending a login request or captured $M_7$ for sending acknowledgement. Let us consider that $A_i$ sends a login request $\{ID_i, SID_i, M_1, M_2, Stat\}$ using captured $M_1$ and $M_2$ to the server $S_i$. Here, $Stat = Login$. The server $S_i$ receives this request and verifies $ID_i$, $SID_i$ and $Stat$. It also generates $M_3$ and tries to match it with $M_1$. Since the received message is a valid login message, it will pass all the verifications. Then, the $S_i$ sets $Stat = Auth$ and sends $\{ID_i, SID_i, M_4, M_5, Stat\}$ to the $A_i$. After receiving $\{ID_i, SID_i, M_4, M_5, Stat\}$ from the $S_i$, he tries to calculate $R_{n3}$ and $M_6$. But, he cannot do that because $X_s$ and $R_{n2}$ are unknown to him. Therefore, he cannot generate $M_7$. If he sends a captured $M_7$ to the server $S_i$, it will fail the verification because the $S_i$ will calculate $M_8$ with recently generated $R_{n3}$ and it will not match with previously generated $R_{n3}$ which was used to calculate $M_7$. Therefore, the $S_i$ will discard the message and terminate the session.

*F. Mutual Authentication*

According to [15], if a scheme is insecure against impersonation attack and server masquerading attack, then it cannot provide mutual authentication. However, we have shown that our scheme can provide security against impersonation attack at subsection VI(*C*) and server masquerading attack at subsection VI(*D*). Therefore, we can say our scheme provides mutual authentication.

*G. Password and Smart Card Recovery*

Our scheme also consists of very good and secure password and smart card recovery options. A user needs to provide a correct and trusted recovery contact during registration and follow the password or smart card recovery steps.

*H. Prevents Denial of Service (DoS) Attack*

If any scheme uses biometric templates directly, then there exists a possibility that sometimes it may fail to match the provided templates with stored templates. It is due to existence of noise or different orientation of imprinting the biometrics, etc. However, our scheme does not use templates directly. We use algorithms which rely on biometric cryptosystem or cancellable biometrics technology and can release unique biometric key from the templates which are relatively close enough. This is how our scheme prevents denial of service (DoS) attack.

*I. Forgery Attack*

In this attack, the attacker $A_i$ may forge into the insecure channel and manage to get messages which are used during login and authentication phase. But, he cannot use these messages to gather any information to generate future login and authentication messages. We have already shown that how the scheme prevents user impersonation attack and server masquerading attack. The attacker may also try to use old messages to gain access. We have shown that how our scheme can resist replay attack. So, considering all these analysis we can say our scheme can resist forgery attack.

*J. Session Key Support*

The session key is required to conduct further secret communication between the user and the server after login. Our scheme provides a mechanism to generate session key during the authentication phase. It reduces the overhead of computation and communication, and also provides the opportunity to conduct further secret communication smoothly.

*K. Comparison with Existing Schemes*

Comparisons among our scheme and few of the existing schemes in terms of security features and functionality are summarized in Table 2.

TABLE 2
SECURITY FEATURES AND FUNCTIONALITY COMPARISON

| Properties | $S_1$ | $S_2$ | $S_3$ | $S_4$ | $S_5$ |
|---|---|---|---|---|---|
| Prevents Password Guessing Attack | Y | Y | N | Y | N |
| Prevents Security Key Stealing | Y | N | N | N | N |
| Prevents User Impersonation Attack | Y | Y | N | Y | N |
| Prevents Server Masquerading Attack | Y | N | N | N | N |
| Prevents Replay Attack | Y | Y | Y | Y | Y |
| Password Recovery | Y | N | N | N | N |
| Smart Card Recovery | Y | N | N | N | N |
| Provides Mutual Authentication | Y | N | N | N | N |
| Prevents Denial of Service Attack | Y | N | N | N | Y |
| Prevents Forgery Attack | Y | N | N | N | Y |
| Supports Session Key | Y | N | N | N | Y |

$S_1$ = Proposed Scheme, $S_2$ = Scheme proposed by Hwang et al. [11], $S_3$ = Scheme proposed by Das [14], $S_4$ = Scheme proposed by An [15], $S_5$ = Scheme proposed by Li et al. [16].
Y = Yes, N = No.

## VII. CONCLUSION

In this paper, we have presented a secure three factor user authentication scheme using biometric and smart card. Through security analysis, we have shown that our scheme outperforms existing schemes in terms of security and features. Our proposed scheme uses the strength of AES to prevent the attacker from stealing data as well as resists several attacks to ensure the security of the login and authentication mechanism. Moreover, it provides password, and smart card recovery options. Our scheme also supports session key agreement to ensure the further secret communication, reduces the overhead of computation and communication. It also uses secure key generation process to generate biometric keys from biometrics. We have depicted a comparison table with few of the existing schemes and our proposed scheme which clearly shows the security advantages of our scheme over those schemes.